\documentclass[twocolumn,twoside,slac_two]{revtex4}
\usepackage{graphicx}
\usepackage{fancyhdr}
\begin{document}

\title{
  Measurements of the CKM angle {\boldmath $\phi_1/\beta $} from Belle and BaBar}

%

\author{K. Vervink, for the Belle Collaboration }
\affiliation{Ecole Polytechnique F\'ed\'erale de Lausanne, Switzerland}

\begin{abstract}
We report recent measurements of the CKM angle $\phi_1$/$\beta$
using large data samples collected by the Belle and BaBar experiments at the
$e^+e^-$ asymmetric-energy colliders.  
\end{abstract}

\maketitle

\thispagestyle{fancy}

\section{Introduction}
In the Standard Model (SM), the irreducible complex phase in the
Cabbibo-Kobayashi-Maskawa (CKM) quark-mixing matrix gives rise to
$CP$ violation~\cite{bib:cabbibo}.  Measurements of the time-dependent
$CP$-asymmetry amplitudes in final states accessible by both $B^0$ and
$\overline{B}^0$ decays probe $\sin 2 \phi_1$, where $\phi_1 = \beta = \arg[V_{cd}
V^*_{cb}]/[V_{td}V^*_{tb}]$ is one of the unitary triangle angles.   The phase
difference $2 \phi_1$ between decays with and without $B^0 -
\overline{B}^0$ mixing arises from box diagrams which mainly occur through a virtual
top quark.  An exclusive measurement of $\sin 2\phi_1$ is possible when no non-trivial relative weak phases appears in the decay
mechanism.

In this article we report measurements obtained by the Belle and BaBar experiments at the
asymmetric-energy $e^+ e^-$ $B$ factories KEK$B$~\cite{bib:kekb} and PEP-II~\cite{bib:pep2}.  Both
accelerators operate at the $\Upsilon(4S)$ resonance, which is
produced with a Lorentz boost of $0.43$ at KEK$B$ and $0.56$ at PEP-II.
At the
time of writing Belle and BaBar collected more than $790\,\rm{fb^{-1}}$
and $550\,\rm{fb^{-1}}$ respectively, which corresponds to a total of approximately $1.4$
billion $B \overline{B}$ events. 

The Belle detector is a large-solid-angle magnetic spectrometer that
consists of a silicon vertex detector (SVD),  a 50-layer central drift
chamber (CDC), an array of Aerogel threshold Cherenkov counters (ACC),
a barrel-like arrangement of time-of-flight scintillation counters
(TOF) and an electromagnetic calorimeter (ECL) comprised of CsI (Ti)
crystals located inside a superconducting solenoid coil that provides
a $1.5\,\rm{T}$ magnetic field.  An iron flux-return located outside the coil
is instrumented to detect $K_L^0$ mesons and to identify muons (KLM).
A detailed description of the Belle detector can be found
elsewhere~\cite{bib:belledetector}. 

The momenta of charged particles
are measured by the BaBar detector with a tracking system consisting of a five-layer
silicon vertex tracker (SVT) and a 40-layer drift chamber (DCH)
surrounded by a $1.5\,\rm{T}$ solenoidal magnet.  An electromagnetic
calorimeter (EMC) comprising $6580$ CsI(Tl) crystals is used to
measure photon energies and positions.  Charged hadrons are
identified with a detector of internally reflected Cherenkov light
(DIRC) and ionization measurements in the tracking detectors. The
BaBar detector is described in detail
elsewhere~\cite{bib:babardetector}. 

\section{Analysis technique}

To measure time-dependent $CP$ asymmetries we typically fully
reconstruct a neutral $B$ meson decaying into a $CP$
 eigenstate.  From the remaining particles in the event, the vertex of
 the other $B$ meson, $B_{\rm{tag}}$, is reconstructed and its
 flavor is identified (tagging).  When assuming  $CP$ conservation in $B^0\overline{B}^0$ mixing
and $\Delta \Gamma /\Gamma = 0$, the time-dependent decay rate of the
 neutral $B$ meson to the $CP$ eigenstate is given by:
\begin{eqnarray}  
\mathcal{P}(\Delta t) &=& \frac{e^{-|\Delta t|/\tau_{B^0}}}{4
  \tau_{B^0}} \Big \{1 + q \Big [\mathcal{S} \sin(\Delta m_d \Delta t) \nonumber \\ 
&+&  \mathcal{A} \cos(\Delta m_d \Delta t) \Big ] \Big \},
\label{eq:basic}
\end{eqnarray}  
where $q = +1 (-1)$ when the other $B$
meson in the event decay is a $B^0$ ($\overline{B}^0$), $\Delta t =
t_{CP} - t_{\rm{tag}}$ is the proper time difference between the two
decays.  $\tau_{B^0}$ is the neutral $B$ lifetime, $\Delta m_d$ the mass
difference between the two $B^0$ mass eigenstates and $\mathcal{S}$ and $\mathcal{A}$  are the $CP$-violating parameters
\begin{eqnarray}  
\mathcal{S} = \frac{2 Im(\lambda)}{|\lambda|^2 + 1}, \quad \mathcal{A}
= \frac{|\lambda|^2 - 1}{|\lambda|^2 + 1}, \nonumber
\end{eqnarray} 
where $\lambda$ is a complex parameter depending on the $B^0-
\overline{B}^0$  mixing as well as on the decay amplitudes 
for both $B^0$ and $\overline{B}^0$ to the $CP$ eigenstate.
Note that in the BaBar convention $\mathcal{A} = -\mathcal{C}$.

When only one diagram contributes to the
decay process and no other weak or strong phases
appear in the process, the SM predicts $\mathcal{A} 
 = 0$ and $\mathcal{S} = - \eta \sin 2
\phi_1$ where $\eta$ is the $CP$ eigenvalue of the final state.  A non-zero value for $\mathcal{A}$ would indicate a direct-$CP$
violation.  Any large measured deviation with respect to 
the prediction can be a sign of New Physics. However
when other diagrams with different weak phases appear in the
interaction, the experimental result of $\mathcal{S}$ will not
necessarily be equal to $ \sin 2 \phi_1$.  The decays presented in
this paper are expected to only have a small deviation from $ \sin 2
\phi_1$ in the SM.

The measurements of $\sin 2 \phi_1$ reported in this paper can be
grouped according to their quark transitions:
\begin{flushleft}
\begin{itemize}
\item [-]  { \boldmath $b \to c\overline{u}d$ }{\bf transitions }: $B^0 \to D_{CP}^{(*)0}
  h^0$;
\item [-] { \boldmath $b \to c\overline{c}s$ }{\bf transitions }: $B^0 \to J/\psi
  K_{\rm{S}}^0$, $B^0 \to J/\psi
  K_{\rm{L}}^0$, $B^0 \to J/\psi K^{*0}$, $B^0 \to \psi(2S) K_{\rm{S}}^0$,
   $B^0 \to \eta_{c} K_{\rm{S}}^0$ and $B^0 \to \chi_{c1}  K_{\rm{S}}^0$;
\item [-] { \boldmath $b \to c\overline{c}d$ }{\bf transitions}: $B^0 \to J/\psi
  \pi^0$ and $B^0 \to D^{*+} D^{*-}$.
\end{itemize}
\end{flushleft}
The decays are grouped from top to bottom with decreasing tree amplitude
or increasing sensitivity to New Physics.  
Finally we will also give an overview of recent measurements of $\cos 2
\phi_1$ provided by $B^0 \to D_{CP}^{(*)0}  h^0$ and  $B^0 \to D^{*+}D^{*-}K_{\rm{S}}^0$ decays.

\section{\boldmath $B^0 \to D_{CP}^{(*)0}  h^0 \,\, (h^0 = \pi^0, \eta, \omega)$}
 
The decay $B^0 \to D_{CP}^{(*)0}  h^0 (h^0 = \pi^0, \eta, \omega)$ is
governed by a color-suppressed $b \to c\overline{u}d$ tree diagram. When the
neutral $D$ meson decays to a $CP$ eigenstate Eq.~\ref{eq:basic}
holds.  The next-to-leading order
  diagram is a doubly Cabbibo and color-suppressed tree diagram with
  the same quark transitions as the main diagram.  Therefore the SM
  corrections on $\sin 2\phi_1$ are expected to be only at the percent level~\cite{bib:dcph0exp}.
However, $R$-parity-violating
super-symmetric processes could enter at the tree level and lead to a
deviation from the SM prediction.  

BaBar reported a measurement of $\sin 2\phi_1$~\cite{bib:dcph0} by
reconstructing the following decay modes  $D^{*0} \to D^0 \pi^0$ and $D^0 \to K^+ K^-$,
$D^0 \to K_{\rm{S}}^0 \pi^0$ and $D^0 \to K_{\rm{S}}^0 \omega$.  The
analysis is performed on $383 \times
  10^6\,B\overline{B}$ pairs of which $340 \pm 32$ signal events are
  reconstructed.  The measured $CP$-violating parameters,
\begin{eqnarray}
\sin 2\phi_1&=&  0.56 \pm 0.23 \,\rm{(stat)} \pm
0.05\,\rm{(syst)} \nonumber \\
\mathcal{A} &=& 0.23 \pm 0.16\,\rm{(stat)} \pm
0.04\,\rm{(syst)} \nonumber, 
\end{eqnarray}
are consistent with the SM expectations.  

\section{\boldmath $B^0 \to c\overline{c}K^0$ transitions}

The $b \to c\overline{c}K^0$ transitions are referred to as the golden modes
due to their relatively large branching fractions $\mathcal{O}(10^{-4}-10^{-5})$, low
experimental background levels and high reconstruction efficiencies.
Typically a signal purity of more than $95\%$ is obtained for $B^0
\to J/\psi(\ell^+\ell^-)K_{\rm{S}}^0(\pi^+ \pi^-)$ decays.  Furthermore the theoretical uncertainties are
small~\cite{bib:golden}.  These modes are dominated by
a color-suppressed $b \to c\overline{c} s$ tree diagram and the dominant
penguin diagram has the same weak phase.  The highest
order term with a different weak phase is a Cabibbo-suppressed penguin
contribution.  Therefore the prediction $\mathcal{S}
= -\sin 2 \phi_1$ and $\mathcal{A} = 0$ is valid to a good accuracy.  Recent theoretical
calculations suggest that the correction on $\mathcal{S}$ is of the
order of $10^{-3}-10^{-4}$~\cite{bib:ccks}.  Because of the high
experimental precision and the low theoretical uncertainty these modes
serve as a benchmark in the SM, which means that any other measurement of $\sin 2
\phi_1$ that has a significant deviation, beyond the usual small SM
corrections, indicates evidence for New Physics.

Both BaBar~\cite{bib:ccksbabar} and Belle~\cite{bib:ccksbelle} studied $CP$ violation in these decays. 
 Belle reconstructed around $7500$ signal events in the  $B^0 \to J/\psi
K_{\rm{S}}^0$ channel and $6500$ signal events in the $B^0 \to J/\psi K_{\rm{L}}^0$ channel using $535 \times
  10^6\,B\overline{B}$ pairs.  BaBar reconstructed additional modes such as
  $J/\psi K^{*0}$, $\psi(2S) K_{\rm{S}}^0$, $\eta_{c} K_{\rm{S}}^0$ and $\chi_{c1}
  K_{\rm{S}}^0$.  Using a data sample of $383 \times
  10^6\,B\overline{B}$ pairs BaBar reconstructed 
  approximately $6900$ $CP$-odd signal events and $3700$ $CP$-even
  signal events.

The results of the time-dependent $CP$ analysis are shown in
Table~\ref{tab:goldenmodes} including the BaBar result using only the
$J/\psi K^0$ modes, to provide a direct comparison with Belle.  The measurements of the
two experiments agree well within the statistical uncertainties.   

Belle recently reported also a measurement of the
$CP$-violation parameters in the $B^0 \to \psi(2S)
K_{\rm{S}}^0$ channel~\cite{bib:psiks}. Using a sample of  $657 \times
  10^6\,B\overline{B}$ pairs, $1284 \pm 38$ signal events are
  reconstructed.  The measured $CP$-violating parameters are included in Table~\ref{tab:goldenmodes}.
\begin{table}[h]
\begin{center}
\caption{$CP$-violating parameters measured by Belle and BaBar with the golden
  modes, the errors are statistical only.}
\begin{tabular}{l l | c | c   }
& & $\sin 2\phi_1$ & $\mathcal{A}$ \\
\hline
BaBar &$J/\psi K^0$ &$  0.697 \pm 0.035$&$ -0.035 \pm 0.025$ \\
BaBar &all $c\overline{c}s$ &$0.714 \pm 0.032$ &$ -0.049 \pm 0.022$ \\
Belle& $J/\psi K^0$ &$  0.642 \pm 0.031$&$\,\,\,\,\, 0.018 \pm 0.021$ \\
Belle &$\psi(2S) K^0_{\rm{S}}$ &$   0.72 \pm 0.09$&$\,\,\,\,\,  0.04 \pm 0.07 $\\
\end{tabular}
\label{tab:goldenmodes}
\end{center}
\end{table}

\begin{figure}[h]
\centering
\includegraphics[width=80mm]{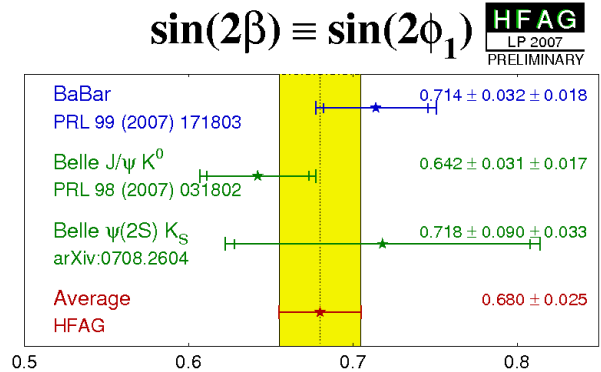}
\caption{Comparison between the Belle and BaBar measurements of $\sin 2 \phi_1$ with $b \to
  c\overline{c}s$ decays.  The bottom line shows the world average.}
\label{fig:ccshfag}
\end{figure}

Figure~\ref{fig:ccshfag} summarizes the results of $\sin 2 \phi_1$ for $b \to
  c\overline{c}s$ decays from Belle and BaBar.  A world average is
  calculated by the Heavy Flavor Averaging Group (HFAG)~\cite{bib:hfag}, 
\begin{eqnarray}
\sin 2\phi_1&=& 0.680 \pm 0.025, \nonumber    
\end{eqnarray}
which reduces the total uncertainty on $\sin 2\phi_1$ to $3.7\%$.   

BaBar also analyzed the $CP$-odd fraction of $b \to c\overline{c}s$ decays containing two vector
  particles.  The $CP$ eigenstate of these decays can be $+1$ or $-1$
  depending on the total angular momentum.  To disentangle the $CP$-odd
  fraction a three-dimensional angular analysis
  is performed on  $232 \times
  10^6\,B\overline{B}$ events~\cite{bib:ccsangul}.  The extracted $CP$-odd fractions are
  shown in Table~\ref{tab:angular}.  The result of $B \to J/\psi
  K^{*}$ is within two standard deviations consistent with $CP$-odd
  fraction measured at the Belle using $277 \times
  10^6\,B\overline{B}$ pairs The $CP$-odd fraction for the neutral $B$
  decay reads $0.195 \pm 0.012\,\rm{(stat)} \pm 0.008\,\rm{(syst)}$ and
  for the charged $B$ decays $0.180 \pm 0.014\,\rm{(stat)} \pm 0.010\,\rm{(syst)}$ ~\cite{bib:angulbelle}.  
\begin{table}[h]
\begin{center}
\caption{$CP$-odd fractions of three vector-vector $b \to c\bar{c}s$ decay modes, measured by BaBar.  The first error mentioned is 
  statistical, the second is systematic.  }
\begin{tabular}{c|c}
Decay & $CP$-odd fraction \\
\hline
$B \to J/\psi K^{*}$& $0.233 \pm 0.010 \pm 0.005 $ \\ 
$B \to \psi(2S) K^{*} $&$ 0.30 \pm 0.06 \pm 0.02 $  \\ 
$B \to \chi_{c1} K^{*}$ &$0.03 \pm 0.04 \pm 0.02$ 
\end{tabular}
\label{tab:angular}
\end{center}
\end{table}

\section{\boldmath $B^0 \to J/\psi \pi^0$}

The $B^0 \to J/\psi \pi^0$ decay takes place through a \mbox{$b \to c\overline{c}d$}
transition.  The dominant tree diagram is Cabibbo suppressed but
contrary to the golden modes, the dominant penguin diagram is of
the same order as the tree diagram and has a different weak phase.  Therefore, even within the SM, the deviation in $\sin 2\phi_1$ could be
substantial. 

Both Belle~\cite{bib:jpsipibelle} and BaBar~\cite{bib:jpsipibabar} have updated their $CP$-violation measurements in
this decay.  Belle performed an analysis on  $535 \times
  10^6\,B\overline{B}$ pairs and obtained $290$ events in the signal region, while BaBar
  used  $466 \times
  10^6\,B\overline{B}$ pairs and obtained $184$ signal events. 
\begin{figure}[h]
\centering
\includegraphics[width=40.mm]{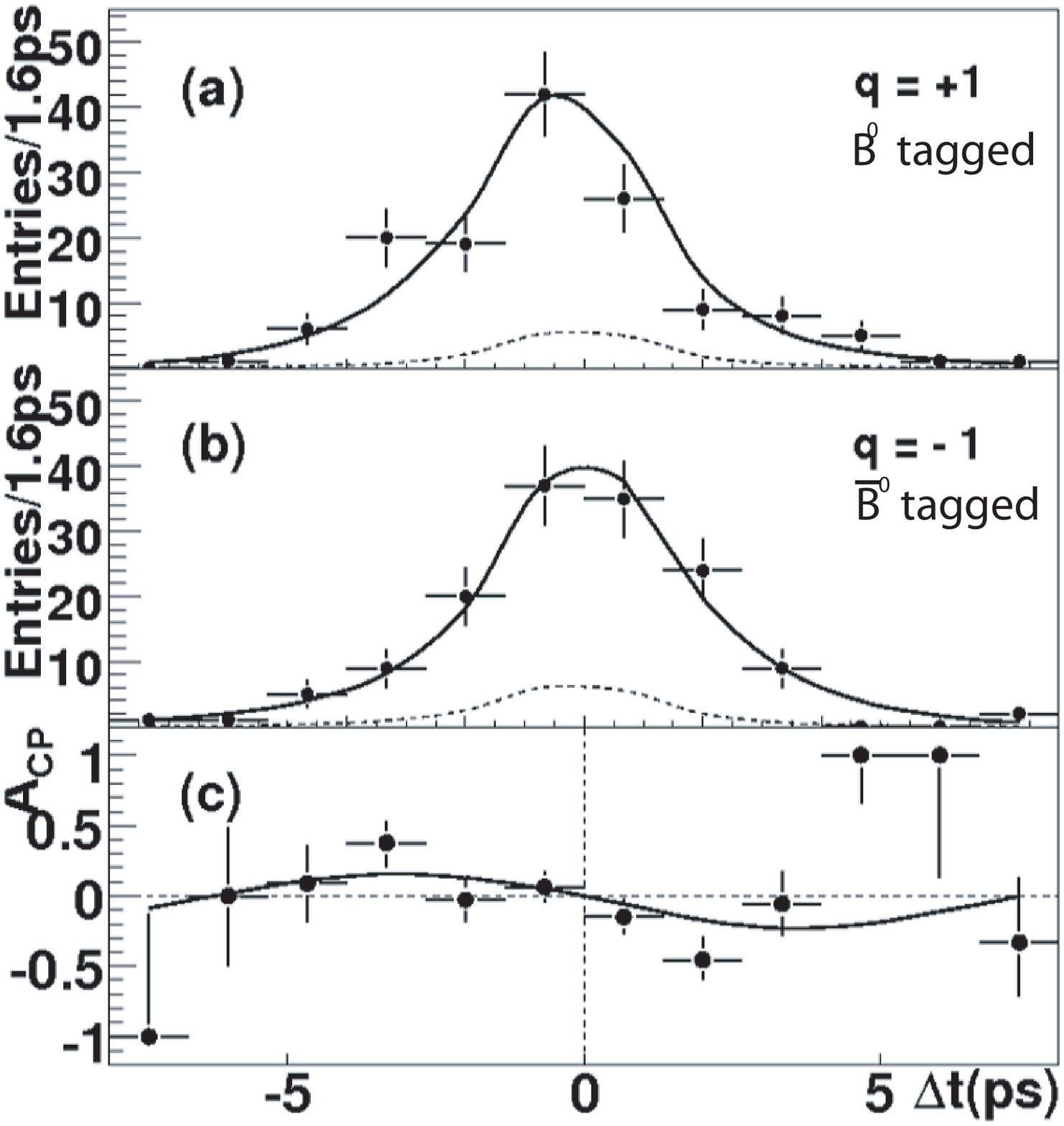}
\includegraphics[width=40.mm]{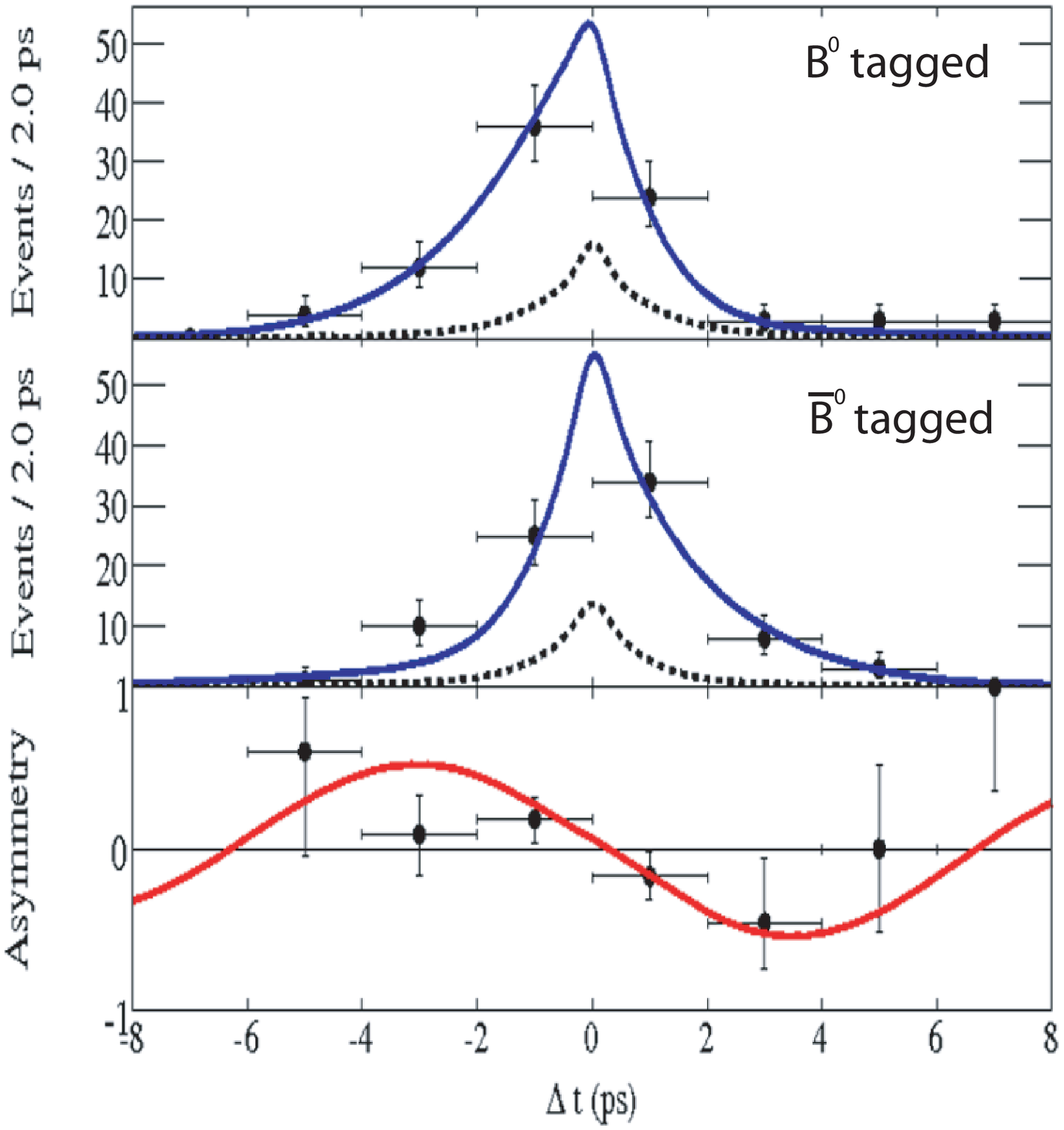}
\caption{The proper time distribution for tagged $B^0$ ($\overline{B}^0$)
  candidates and the raw asymmetry of $B^0 \to J/\psi \pi^0$ as a
  function of $\Delta t$ measured by  BaBar (right) and Belle (left).  The lines represent the fit results.  }
\label{fig:jpsipi}
\end{figure}
The plots in Figure~\ref{fig:jpsipi} show the proper time distribution and the
raw asymmetry, defined as  $(N_{+} - N_{-})/(N_{+} +
N_{-})$, where $N_{+} (N_{-})$ is the number of observed neutral $B$ candidates
with $q = +1 (-1)$.  Within the experimental uncertainties, the results
are compatible with the SM prediction.  The measured $CP$ parameters are summarized in Table~\ref{tab:jpsipi}.
\begin{table}[h]
\begin{center}
\caption{$CP$-violating parameters in the $B^0 \to J/\psi \pi^0$ decay
  measured by Belle and BaBar.  The first error mentioned is the
  statistical, the second is systematic.   }
\begin{tabular}{l | c | c   }
& Belle & BaBar \\
\hline
$\sin 2\phi_1$&$  0.65 \pm 0.21 \pm
0.05$&$  1.23 \pm 0.21 \pm
0.04$ \\
$\mathcal{A}$ &$0.08 \pm 0.16\pm
0.05$&$  0.20 \pm 0.19  \pm
0.03 $
\end{tabular}
\label{tab:jpsipi}
\end{center}
\end{table}
The BaBar result shows an evidence of $CP$ violation, obtained with a $4 \sigma$ significance.

\section{\boldmath $B^0 \to D^{*+} D^{*-}$}
\begin{figure}[h]
\centering
\includegraphics[width=80mm]{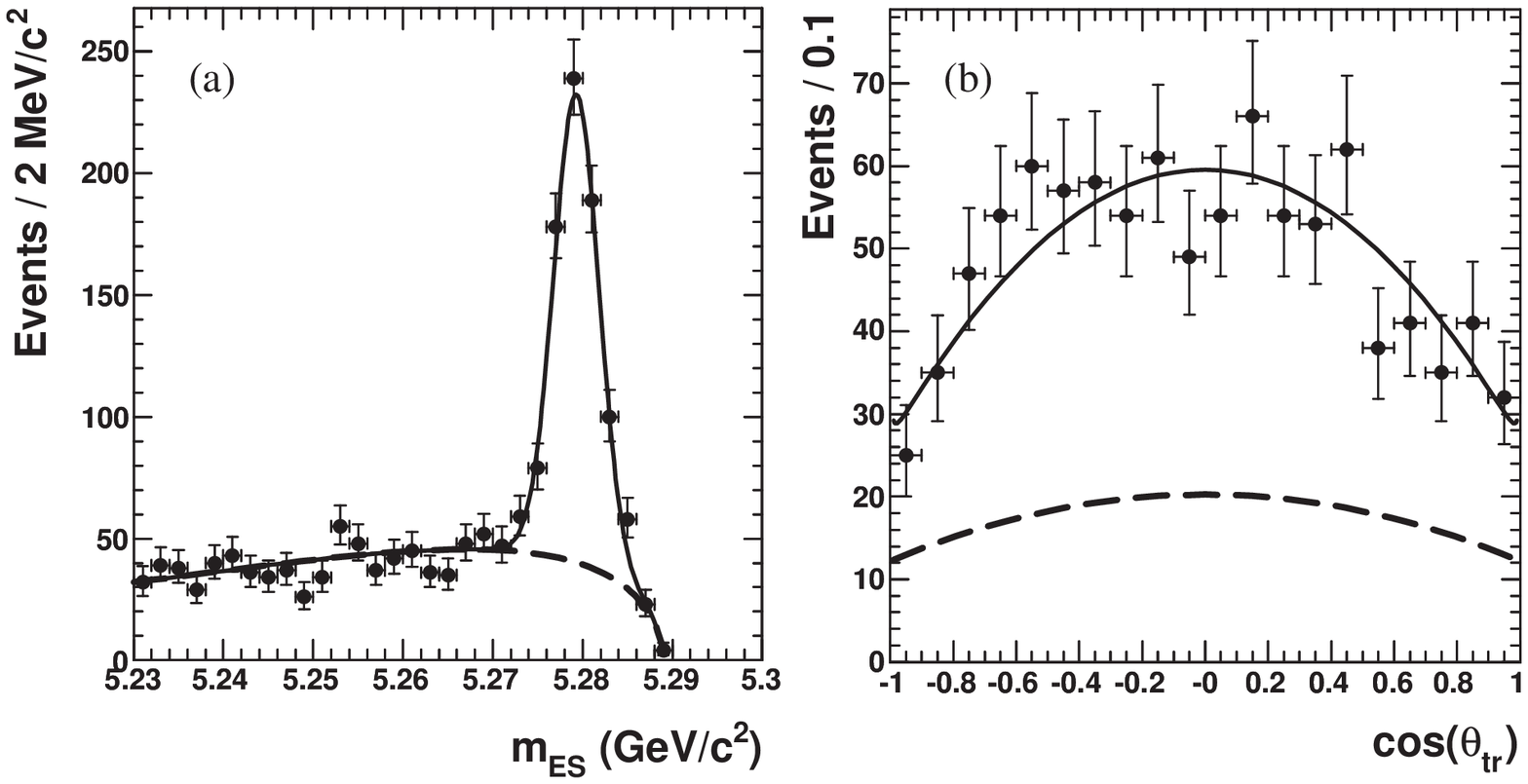}
\includegraphics[width=40.5mm]{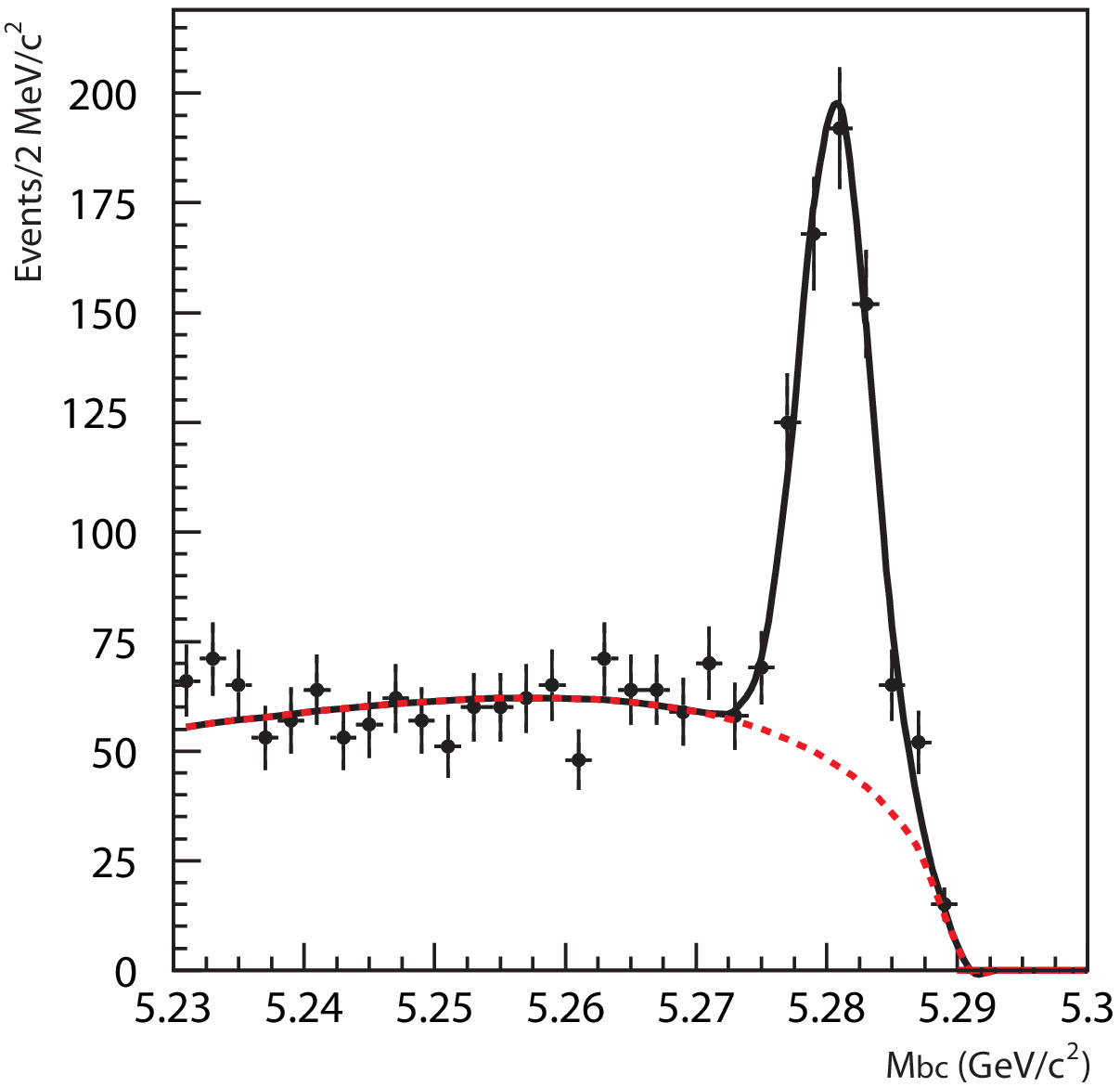}
\includegraphics[width=39.5mm]{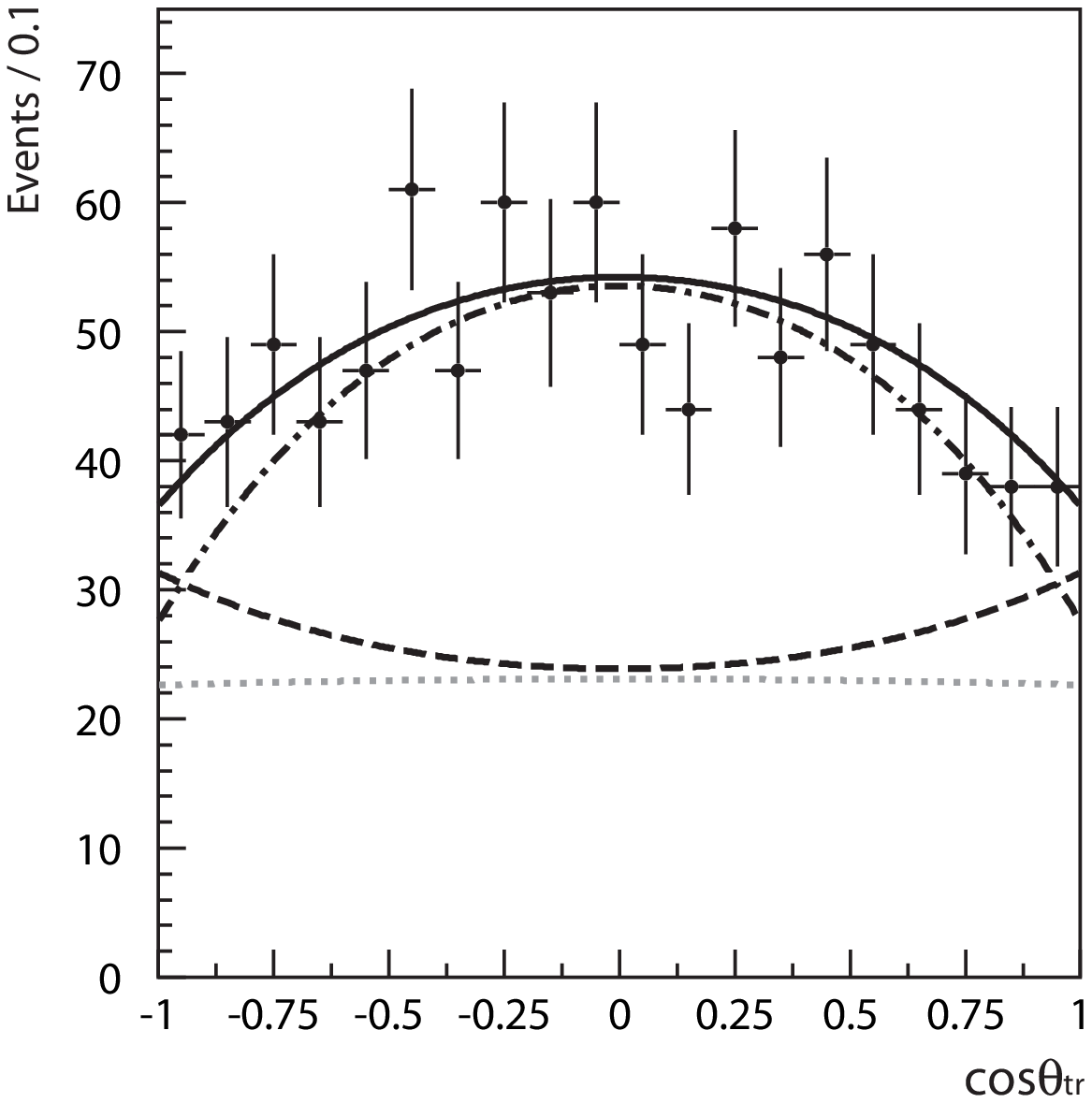}
\caption{The measured $M_{\rm{bc}}$ and $\cos \theta_{\rm{tr}}$
  distributions in the signal region for BaBar (top) and Belle
  (bottom).  The solid lines represent the projections of the fit
  results, the dotted lines are the background components.  }
\label{fig:dstdstsignal}
\end{figure}

The measurements of the double charm decay $B^0 \to D^{*+} D^{*-}$ are
updated by both experiments~\cite{bib:dstdstbabar} and the high statistics signals are shown
in the left plots of Figure~\ref{fig:dstdstsignal}. The Belle
analysis is performed on  $657 \times
  10^6\,B\overline{B}$ pairs and has extracted $545 \pm 29$ signal events while
  the BaBar analysis found $638 \pm 38$ signal events in $383 \times
  10^6\,B\overline{B}$ pairs
The tree amplitude is CKM-suppressed and the
contribution of penguin diagrams in this decay is estimated to be at
the percent level~\cite{bib:dstdst}.  The $CP$ eigenvalue of the
$D^{*+} D^{*-}$ pair is $+1$ when it decays via a $S$ and $D$ wave or $-1$
for a $P$ wave.  A helicity 
study is performed to extract the $CP$-odd fraction, $R_{\rm{odd}}$, which dilutes the measurement of $\mathcal{S}$.
The angular analysis is performed in the so-called transversity basis
and the Belle and BaBar results are shown in the right plots of Figure~\ref{fig:dstdstsignal}. The
extracted $CP$-odd fraction is $R_{\rm{odd}} = 0.143 \pm 0.034\,\rm{(stat)} \pm
0.008\,\rm{(syst)}$ for BaBar and $R_{\rm{odd}} = 0.116 \pm 0.042\,\rm{(stat)} \pm
0.004\,\rm{(syst)}$ for Belle, consistent with previous
measurements.  

\begin{figure}[h]
\centering
\includegraphics[width=40mm]{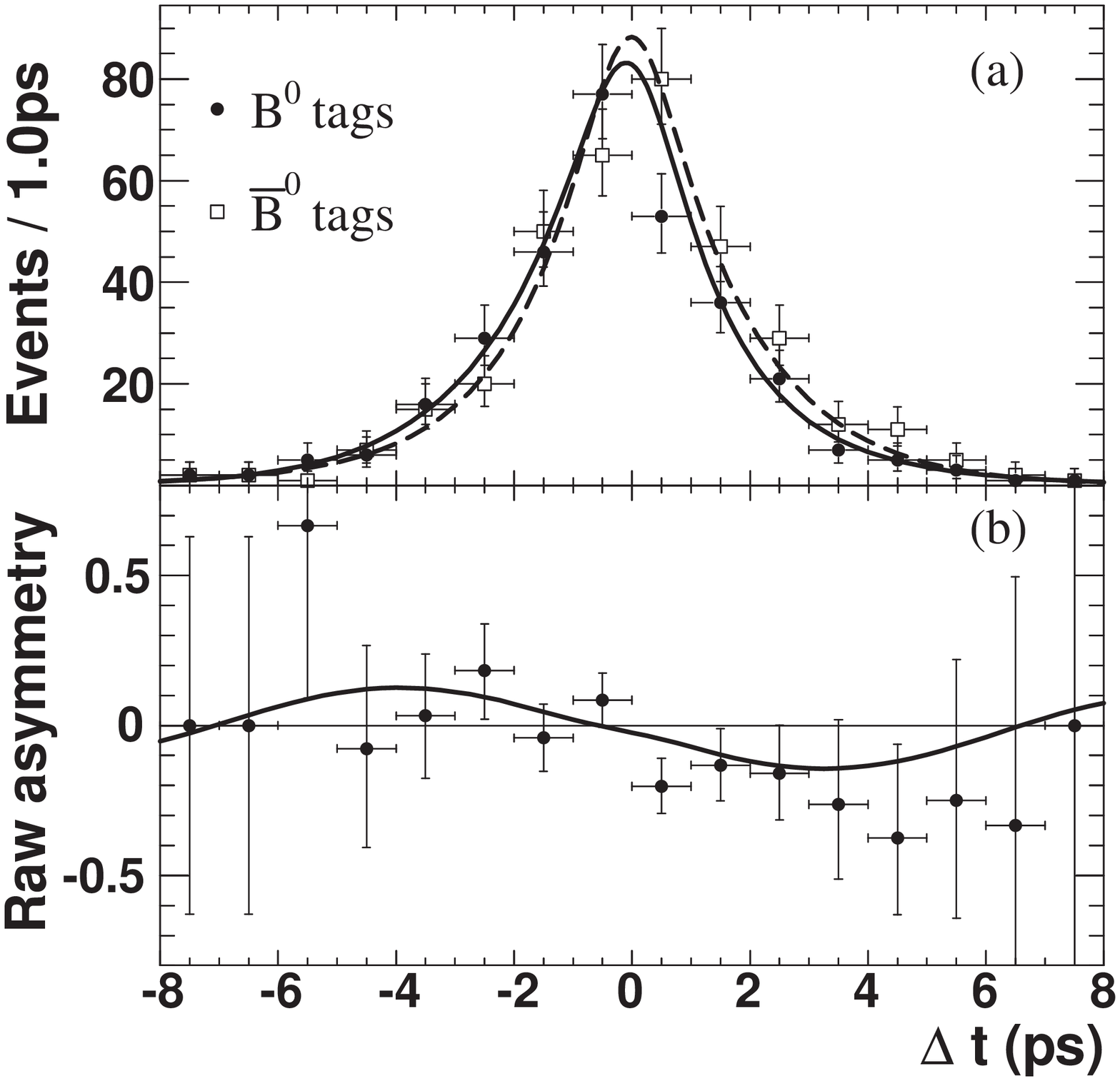}
\includegraphics[width=40mm]{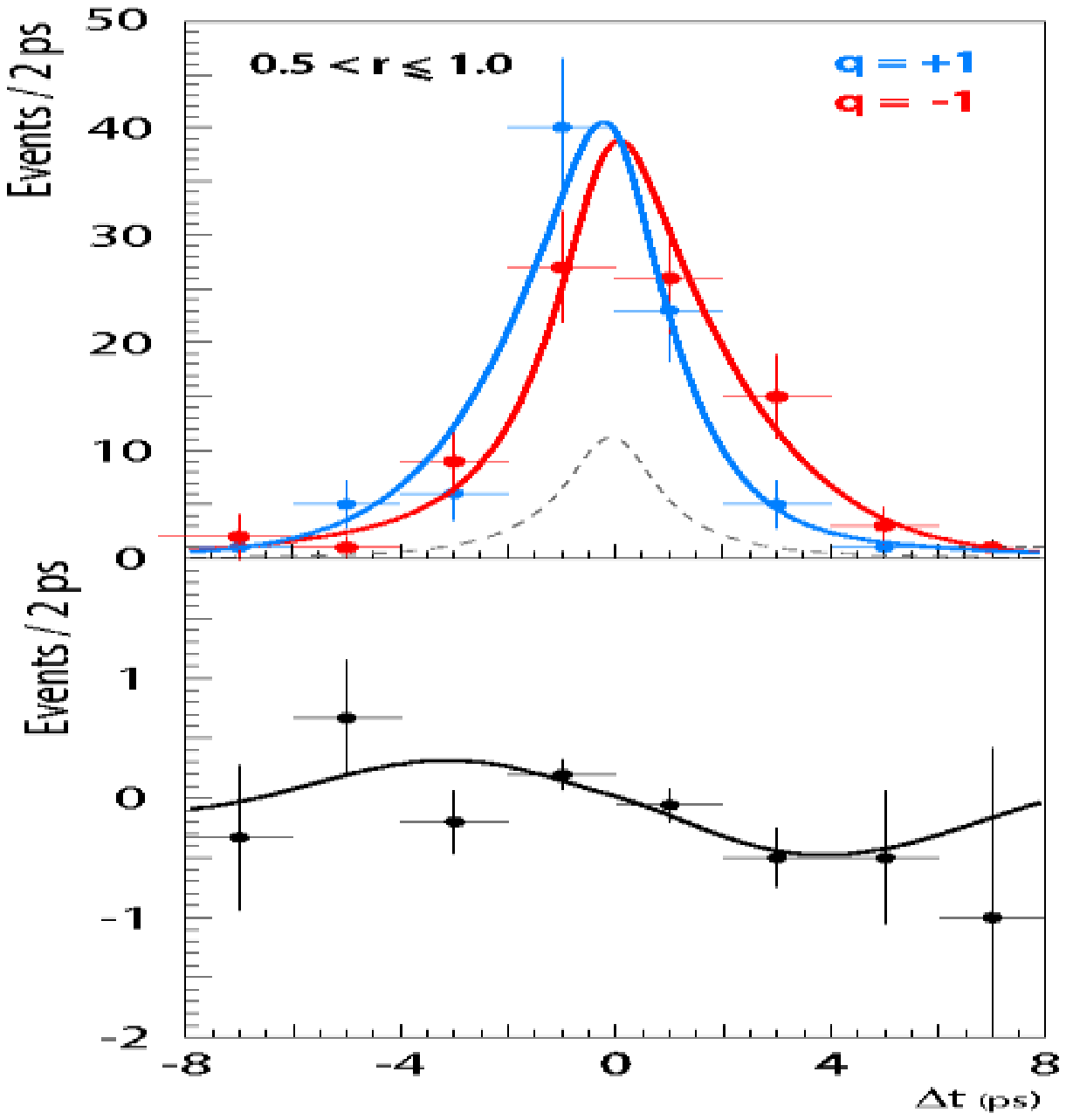}
\caption{Top: the $\Delta t$ distributions of $B^0 \to D^{*+} D^{*-}$
  events in the signal region for $B^0$ ($\overline{B}^0$) tagged
  candidates. Bottom: the raw asymmetry as a function of $\Delta t$.
  The lines represent the fit result. The plots on the left show the
  BaBar result while the plots on the right are the
  Belle results.}
\label{fig:dstdstcp}
\end{figure}

Figure~\ref{fig:dstdstcp} shows the  $\Delta t$ distribution and the raw asymmetry for events with
a good-quality tag. The fitted $CP$ parameters are consistent with
each other and the SM predictions.  BaBar found $\mathcal{A} = 0.02 \pm
0.11\,\rm{(stat)}\pm 0.02\,\rm{(syst)}$ and $\sin 2 \phi_1 =
0.66 \pm 0.19\,\rm{(stat)}\pm 0.04\,\rm{(syst)}$ while the Belle
result is: $\mathcal{A} = 0.16 \pm
0.13\,\rm{(stat)}\pm 0.02\,\rm{(syst)}$ and $\sin 2 \phi_1 =
0.93 \pm 0.24\,\rm{(stat)}\pm 0.15\,\rm{(syst)}$.
\section{Measurement of \boldmath $\cos 2 \phi_1$}
\label{sec:cosphi1}

The measurements of $\sin 2 \phi_1$ leaves a 4-fold ambiguity in the
value of $\phi_1$ which can be partially resolved by measuring
$\cos 2 \phi_1$.  We will show two analyzes in this section that
measured the sign $\cos 2 \phi_1$.

In the first analysis both $B^0$ and $\overline{B}^0$ mesons decay to the final state
$D^{*+}D^{*-}K_{\rm{S}}^0$.  A potential interference effect of the
decay proceeding though an intermediate resonance can be measured by
dividing the $B$-decay Dalitz plot into regions with \mbox{$ s^+ >(<)\,s^-$}, where $s^{\pm} \equiv m^2 (D^{*\pm}
K_{\rm{S}}^0)$~\cite{bib:dstdstk}.  Belle performed a measurement on  $449 \times
  10^6\,B\overline{B}$ pairs which corresponds to $131.2^{+ 14.8}_{-
    14.1}$ extracted signal events and measured $2J_{\rm{s2}}/J_0 \cos
  \phi_1 = -0.23^{+0.43}_{-0.41} \,\rm{(stat)} \pm
  0.13\,\rm{(syst)}$~\cite{bib:dstdstkbelle}, where $J_{\rm{s2}}$ and $J_{\rm{0}}$ are the integrals
  over the half-Dalitz space, $s^- > s^+$ of $|a|^2 - |\overline{a}|^2$ and
  the imaginary component, $Im(\overline{a}a^*)$ respectively, where
  $a(\overline{a})$ are the decay amplitudes of $B^0 (\overline{B}^0)
  \to D^{*+} D^{*-} K_{\rm{S}^0}$.  Although the sign of the factor
  $2J_{s2}/J_{0}$ can be deduced from theory, a model-independent sign of $\cos 2
  \phi_1$ could not be obtained given the errors.  A similar analysis is also performed by BaBar on  $230 \times
  10^6\,B\overline{B}$ pairs and concluded $\cos 2
  \phi_1 > 0$ with $94\%$ confidence level~\cite{bib:dstdstkbabar}.  

A second technique to determine the sign of  $\cos 2
\phi_1 $ utilizes the decay $B^0 \to D^{*} (\pi^+ \pi^-
K_{\rm{S}}^0)h^0$, where $h^0 = \eta, \eta', \pi^0$ or $\omega$.  This
decay can occur with and without $B^0\overline{B}^0$
mixing and interference effects are visible across the $D^0$ Dalitz
plot.  Belle performed this analysis on $386 \times
  10^6\,B\overline{B}$ events~\cite{bib:dalitzdhbelle} and the fit of the full Dalitz plot gives $\sin 2
\phi_1 = 0.78 \pm 0.44\,\rm{(stat)}\pm 0.22 \rm{(syst + model)}$ and
$\cos 2\phi_1 = 1.87^{+
  0.40}_{-0.53}\,\rm{(stat)}^{+0.22}_{-0.32}\,\rm{(syst +
  model)}$, which gives a preferred positive sign of $\cos 2\phi_1$ at
$96.8 \%$ confidence level.  The result from
BaBar~\cite{bib:dalitzdhbabar} on  $383 \times
  10^6\,B\overline{B}$ events reads $\sin 2
\phi_1 = 0.29 \pm 0.34\,\rm{(stat)}\pm 0.03\,rm{(syst)}\pm 0.05\,\rm{(model)}$ and
$\cos 2\phi_1 = 0.42 \pm 0.49\,\rm{(stat)} \pm 0.09\,\rm{(syst)}\pm
0.13\,\rm{(model)}$ leading to a preferred positive sign for $\cos 2\phi_1$ at
$86 \%$ confidence level.

\section{Conclusions}
Various decay modes have been used by Belle and BaBar to measure $\sin
2 \phi_1$ using high statistics $B\overline{B}$ samples.  The two
experiments have also performed measurements
of the sign of $\cos 2 \phi_1$, which is preferred to be positive.
The measurements of the $CP$-violating parameters in the $b\to
c\overline{c}s$ channels are the most precise results available and
given the positive sign of $\cos 2 \phi_1$, the world average gives
$\phi_1 = 21.5^o \pm 1.0^o $ and $\phi_1 = 201.5^o \pm 1.0^o $, where
the first is conform with the Standard Model.   Finally, the new results of $B$ decays to $ D_{CP}^{(*)0}  h^0
\,\, (h^0 = \pi^0, \eta, \omega)$, $J/\psi \pi^0$ and
$D^{*+} D^{*-}$ are shown.  The $CP$ violating parameters are consistent with the
Standard Model expectations within the uncertainties of the
measurement.


\bigskip 

\end{document}